\documentclass[12pt]{article}
\usepackage[T2A]{fontenc}
\usepackage[utf8]{inputenc}
\usepackage{setspace,amsmath}
\usepackage{bbold}
\usepackage{amsfonts}
\usepackage{amssymb}
\usepackage{amsthm}
\usepackage{cite}
\usepackage{fontenc}
\usepackage[unicode,linktocpage=true,plainpages=false,pdfpagelabels=false]{hyperref}
\usepackage{bm}
\usepackage{float}
\usepackage{feynmp-auto}
\usepackage[dvips]{graphicx}
\usepackage{epstopdf}
\usepackage{authblk}
\usepackage{lipsum}

\usepackage{cancel}
\usepackage{calligra}

\begin{document}
\begin{center}
		{\Large\textbf{Two-Loop Cutoff Renormalization of 4-D Yang--Mills Effective Action}}
		\vspace{0.5cm}
		
		{\large A.~V.~Ivanov$^\dag$ and N.~V.~Kharuk$^\ddag$}
		
		\vspace{0.5cm}
		
		$^\dag${\it St. Petersburg Department of Steklov Mathematical Institute of
			Russian Academy of Sciences,}\\{\it 27 Fontanka, St. Petersburg 191023, Russia}\\
		$^\dag${\it Euler International Mathematical Institute, 10 Pesochnaya nab.,}\\
		{\it St. Petersburg 197022, Russia}\\
		{\it E-mail: regul1@mail.ru}\\
		$^\ddag${\it ITMO University, St.Petersburg 197101, Russia}\\
		{\it E-mail: natakharuk@mail.ru}	
\end{center}
\vskip 10mm
\date{\vskip 20mm}
\vskip 10mm
\begin{abstract}
	In the paper we study the Yang--Mills effective action in the four-dimensional space-time by using background field formalism. We give an explicit way of cutoff regularization procedure, then do a two-loop renormalization and calculate a second $\beta$-function coefficient. We also show that the two-loop singularity contains only logarithmic part in the first degree. At the same time additional properties of a Green function regular part are obtained.
\end{abstract}

\newpage

\tableofcontents  
\newpage

\section{Introduction}
The Yang--Mills fields were introduced in the paper \cite{1}. Since then they have found their geometrical \cite{2} and physical \cite{8} meanings. One of the most fruitful ways to study these fields is a path integral formulation (see \cite{3}), which allows you to consider quantum corrections by loop decomposition (Feynman diagrams, see \cite{4}). However, this approach faces some problems related to divergent integrals, so the way requires some improvement.\\

Renormalization theory \cite{6,7} makes it possible to eliminate \textquotedblleft bad\textquotedblright\,terms from the Feynman diagrams, while the Yang--Mills theory becomes finite and gives accurate results. Nevertheless, this procedure can be used with various regularizations \cite{9,10}, the most widespread of which are dimensional \cite{5,19}, cutoff, and regularization by higher covariant derivatives \cite{w1,w2}. The first and the third cases make it quite easy to produce multi-loop calculations \cite{11,12,13,14,15,18,20}, while for the second one only the first correction \cite{16,17} is known. We also want to note some other works \cite{w5,w6,w7,w8} in which the cutoff regularization was studied earlier. However, they have significant differences and are not related to this paper.\\

The Yang--Mills theory in the four-dimensional space-time has such features as asymptotic freedom and dimensional transmutation. Because of it recent papers (see \cite{21,22}) have described a renormalization \textquotedblleft scenario\textquotedblright\,\,for the Yang--Mills theory with the cutoff regularization. Such approach is convenient due to the presence of the Gell-Mann--Low equation for a coupling constant. However, the works do not contain a procedure of regularization and any explicit calculations.\\

In the present work we give a clear way to regularize and study the two-loop cutoff renormalization of the Yang--Mills effective action. We also demonstrate that the second correction has a singularity only of logarithmic type in the first degree and compute the second $\beta$-function coefficient.\\

\textbf{Structure:} The work consists of several parts. In Section \ref{s2} we give the necessary basic information on the Yang--Mills theory and related methods. Then, in Section \ref{s3}, the results are formulated. Later Section \ref{s4} contains a Green function expansion and a cutoff regularization procedure. After that in Sections \ref{s5} and \ref{s6} we calculate one-loop and two-loop corrections.

\section{Yang--Mills effective action}
\label{s2}
\subsection{Classical action}

To describe the Yang--Mills theory in the four-dimensional space-time we need to introduce some basic concepts. Without loss of generality we work with the Euclidean analog of the theory (not Minkowski space), so the metric tensor is $\delta^{\mu\nu}$. Here and further the Greek letters $\alpha,\beta,\mu,\nu\ldots$ are used to indicate the space indices. Then let $G$ be a compact semisimple Lie group, and $\mathfrak{g}$ is its Lie algebra. Let $t^a$ be the generators of the algebra $\mathfrak{g}$, where $a = 1, \ldots,\dim \mathfrak{g}$, such that the relations hold
\begin{equation}
[t^a,t^b]=f^{abc}t^c,\,\,\,\,\,\,
\mathrm{tr}(t^at^b)=-2\delta^{ab},
\end{equation}
where $f^{abc}$ are antisymmetric structure constants for $\mathfrak{g}$, and $"\mathrm{tr}"$ is the Killing form. We work with an adjoint representation, so it is easy to verify that the structure constants have properties
\begin{equation}
\label{eq4}
f^{cka}f^{dka}=c_2\delta^{cd},\,\,\,\,\,\,
f^{kca}f^{aed}f^{dgk}=-\frac{c_2}{2}f^{ceg},
\end{equation}
where $c_2$ is a normalization constant for the Lie group $G$.\\

Then let $A_\mu=A^a_\mu t^a$ be a smooth Yang--Mills field, and $\hat{F}_{\mu\nu}=\hat{F}_{\mu\nu}^at^a$ is the field strength tensor, components of which can be written in the form
\begin{equation*}
\hat{F}_{\mu\nu}^a=\partial_\mu^{}A_\nu^a-\partial_\nu^{}A_\mu^a+f^{abc}A_\mu^bA_\nu^c.
\end{equation*} 

Under the conditions described above we can define a classical Yang--Mills action (see \cite{3}) as
\begin{equation}
S[A]=\frac{1}{4g^2}\int_{\mathbb{R}^4}d^4x\,\hat{F}^a_{\mu\nu}\hat{F}^{a}_{\mu\nu},
\end{equation} 
where $g=\sqrt{\alpha}/2$ is a coupling constant. Sometimes we use $\alpha$ instead of $g$ because of the connection with the paper \cite{22}.

\subsection{Path integral formulation}
First of all, we need to introduce an object $W$ as a path integral
\begin{equation}
\label{eq23}
e^{-W}=\int_{H}\mathcal{D}A\,e^{-S[A]},
\end{equation}
where $H$ is a functional set, which is determined by physical reasons and asymptotic behaviour of the fields $A_\mu$ at infinity \cite{23}. Then we use a background field method in the form \cite{24,25,26}. It means that we do the following shift
\begin{equation*}
A_\mu=B_\mu+ga_\mu,
\end{equation*}
where the background field components $B_{\mu}$ satisfy the asymptotic behaviour at infinity and a quantum equation of motion, which is just a sum of classical equation of motion and one-particle irreducible ($1\mathrm{PI}$) quantum corrections (see \cite{23,34}). We assume that the components $B_\mu$ have the same gauge transformation rule as the $A_\mu$. Let us define a derivative in the form $D_\mu=\partial_\mu+B_\mu$,
where the component $B_\mu$ acts by using the adjoint representation. Thereby we can rewrite the derivative in the matrix form as
\begin{equation*}
D_\mu^{ab}=\partial_\mu^{}\delta^{ab}+f^{acb}B_\mu^{\,c}.
\end{equation*} 
Then, defining two objects
\begin{equation*}
F_{\mu\nu}=\hat{F}_{\mu\nu}|_{A=B},\,\,\,\,\,\,
W_{-1}=4g^2S[B],
\end{equation*} 
we have the following decomposition of the classical action
\begin{align*}
S[B+ga]=\frac{1}{4g^2}W_{-1}+\frac{1}{g}\Gamma_1(a)+\frac{1}{2}\int_{\mathbb{R}^4}d^4x\,&a_\mu^a M_{1\mu\nu}^{\,\,\,ab}\,a_\nu^b\\
&+g\Gamma_3(a)+g^2\Gamma_4(a)-\frac{1}{2}\int_{\mathbb{R}^4}d^4x\left(D_\mu^{ab} a_\mu^{\,b}\right)^2,
\end{align*}
where
\begin{equation*}
\Gamma_1(a)=-\int_{\mathbb{R}^4}d^4x\,a_\nu^{\,a}D_\mu^{ab}F_{\mu\nu}^b,\,\,\,\,\,\,
\Gamma_3(a)=\int_{\mathbb{R}^4}d^4x\,D_\mu^{ae}a_\nu^{\,e}f^{abc}a_\mu^{\,b}a_\nu^{\,c},
\end{equation*}
\begin{equation*}
\Gamma_4(a)=\frac{1}{4}\int_{\mathbb{R}^4}d^4x\,f^{abc}a_\mu^{\,b}a_\nu^{\,c}f^{aed}a_\mu^{\,e}a_\nu^{\,d},\,\,\,\,\,\,
M_{1\mu\nu}^{\,\,\,ab}=-D_\rho^{ac}D_\rho^{cb}\delta_{\mu\nu}^{}-2f^{acb}F_{\mu\nu}^c .
\end{equation*}
Then, introducing ghost fields $c$ and $\bar{c}$ (see \cite{27}), we can provide a gauge-fixing term and Faddeev--Popov term in the following form
\begin{equation*}
\frac{1}{2}\int_{\mathbb{R}^4}d^4x\left(D_\mu^{ab} a_\mu^{\,b}\right)^2+
\int_{\mathbb{R}^4}d^4x\,\bar{c}^{\,a}M_0^{ab}c^{\,b}+g\,\Omega_3(a,c,\bar{c}),
\end{equation*}
where
\begin{equation*}
M_0^{ab}=-D_\mu^{ae}D_\mu^{eb},\,\,\,\,\,\,
\Omega_3(a,c,\bar{c})=\int_{\mathbb{R}^4}d^4x\,D_\mu^{ab}c^{\,b}f^{aed}a_\mu^{\,e}\bar{c}^{\,d}.
\end{equation*}
It means that we have the following expression for the formula (\ref{eq23})
\begin{align}
\label{eq24}
e^{-W}=e^{-S[B]}\int_{H_0}\mathcal{D}a\,\mathcal{D}c\,\mathcal{D}\bar{c}\,\exp\bigg\{
&-\frac{1}{2}\int_{\mathbb{R}^4}d^4x\,a_\mu^a M_{1\mu\nu}^{\,\,\,ab}\,a_\nu^b
-\int_{\mathbb{R}^4}d^4x\,\bar{c}^{\,a}M_0^{ab}c^{\,b}
\\\nonumber
&-\frac{1}{g}\Gamma_1(a)-g\Gamma_3(a)-g^2\Gamma_4(a)-g\,\Omega_3(a,c,\bar{c})\bigg\},
\end{align}
where, actually, $W=W[B]$ is a functional of the fields $B_{\mu}$, and the integration measure is normalized according to the formula
\begin{equation*}
\int_{H_0}\mathcal{D}a\,\mathcal{D}c\,\mathcal{D}\bar{c}\,\exp\bigg\{
-\frac{1}{2}\int_{\mathbb{R}^4}d^4x\,a_\mu^a M_{1\mu\nu}^{\,\,\,ab}\,a_\nu^b
-\int_{\mathbb{R}^4}d^4x\,\bar{c}^{\,a}M_0^{ab}c^{\,b}\bigg\}=
\frac{\det(M_0/M_0|_{B=0})}{\sqrt{\det(M_1/M_1|_{B=0})}}.
\end{equation*}

\subsection{Perturbation theory}
\begin{figure}[H]
\centerline{\includegraphics[width=0.45\linewidth]{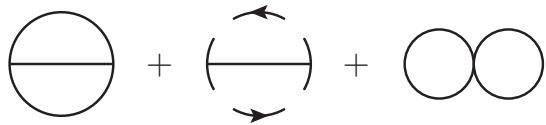}}
\caption{Two-loop Feynman diagrams.}
\label{loop}
\end{figure}

We assume that the coupling constant $g$ is small enough. So we can use a decomposition of the exponentials in a series. For that purpose let us introduce the unit
\begin{equation}
1=\left.e^{Ja+\bar{b}c+b\bar{c}}\right|_{J=\bar{b}=b=0}
\end{equation}
into the integral (\ref{eq24}). Thus we take vertices $\Gamma_1$, $\Gamma_3$, $\Gamma_4$, and $\Omega_3$ out of the integral as functional derivatives
\begin{equation*}
\exp\bigg\{-\frac{1}{g}\Gamma_1\left(\frac{\delta}{\delta J}\right)-g\Gamma_3\left(\frac{\delta}{\delta J}\right)-g^2\Gamma_4\left(\frac{\delta}{\delta J}\right)-g\,\Omega_3\left(\frac{\delta}{\delta J},\frac{\delta}{\delta\bar{b}},\frac{\delta}{\delta b}\right)\bigg\}.
\end{equation*}
Then let us define Green functions $G_0$ and $G_1$ for the Laplace-type operators $M_0$ and $M_1$ by the equalities
\begin{equation}
\label{green}
M_{1\mu\nu}^{\,\,\,ab}G_{1\nu\rho}^{\,\,\,bc}(x,y)=\delta^{ac}\delta_{\mu\rho}\delta(x-y),\,\,\,\,\,\,
M_0^{ab}G_{0}^{bc}(x,y)=\delta^{ac}\delta(x-y).
\end{equation}
Thereby the remaining integrals are Gaussian and can be calculated explicitly
\begin{align*}
\frac{\det(M_0/M_0|_{B=0})}{\sqrt{\det(M_1/M_1|_{B=0})}}
\exp\bigg\{
\frac{1}{2}\int_{\mathbb{R}^4}d^4x\int_{\mathbb{R}^4}d^4y\,&J_\mu^a(x) G_{1\mu\nu}^{\,\,\,ab}(x,y)\,J_\nu^b(y)
\\&+\int_{\mathbb{R}^4}d^4x\int_{\mathbb{R}^4}d^4y\,b^{\,a}(x)G_0^{ab}(x,y)\bar{b}^{\,b}(y)\bigg\},
\end{align*}
where the exponential is called a generating functional and is denoted by $Z[J,b,\bar{b}]$. Hence we can use the perturbation theory for the functional $W[B]$ and decompose it in a series in the powers of the coupling constant $g$ in the following form
\begin{align}
\label{eq20}
W[B]&=\frac{1}{4g^2}W_{-1}+\bigg\{\frac{1}{2}\ln\det(M_1/M_1|_{B=0})-\ln\det(M_0/M_0|_{B=0})\bigg\}\\\nonumber
&-g^2\left.\bigg\{
\frac{1}{2}\Gamma_3^2\left(\frac{\delta}{\delta J}\right)+\frac{1}{2}\Omega_3^2\left(\frac{\delta}{\delta J},\frac{\delta}{\delta\bar{b}},\frac{\delta}{\delta b}\right)
-\Gamma_4\left(\frac{\delta}{\delta J}\right)\bigg\}
Z[J,b,\bar{b}]\right|_{J=b=\bar{b}=0}^{only\,\,\mathrm{1PI}\,\,part}\\\nonumber
&+O(g^4).
\end{align}
Using the diagram technique language \cite{23} we can represent the second line of the last formula as it is depicted on the Figure \ref{loop}. However we do not use diagrams in the work, because such formalism is not very useful for our approach.
\noindent Now we are ready to define an \textbf{effective action} as
\begin{equation}
\label{eq46}
W_{eff}[B]=W[B]-W[0].
\end{equation}
In this case, the main term is equal to the classical action and is proportional to $g^{-2}$. Then we write the first two quantum (one-loop and two-loop) corrections which are proportional to $g^0$ and $g^2$ respectively.

\section{Results}
\label{s3}
According to the renormalization \textquotedblleft scenario\textquotedblright\cite{22} we need to introduce the cutoff regularization (see Section \ref{cutoff}). After that we obtain a new parameter $\Lambda$ and a set of singularities. So we rewrite the effective action (\ref{eq46}) in the form
\begin{equation}
W_{eff}^{\Lambda}[B]=\frac{1}{\alpha}W_{-1}+\underset{one-loop\,\,term}{\big(W_0^{fin}+W_0^{div}\big)}+\alpha\underset{two-loop\,\,term}{\big(W_1^{fin}+W_1^{div}\big)}+O(\alpha^2),
\end{equation}
where the corrections were split into finite and divergent parts. Of course, the corrections are functionals of $B_{\mu}$ and $\Lambda$. The one-loop singularity is a well known result and has a form (see Section \ref{s5})
\begin{equation}
W_0^{div}=-\frac{11L}{6}\frac{c_2}{(4\pi)^2}W_{-1},
\end{equation}
where $L=\ln(\Lambda/\mu)$. The second loop is calculated in this paper and, as shown in Section \ref{s6}, is equal to the singular part of the six summands $\sum_{i=1}^{6}(\mathcal{J}_i\big|_{B=0}-\mathcal{J}_i)$, which are defined by formulas (\ref{eq41}-\ref{eq44}), (\ref{eq45}), and (\ref{eq40}), and is computed in the formula (\ref{eq57}). Under the conditions described above the main result is
\begin{equation}
\label{eq58}
W_1 ^{div}=-\frac{9L}{8}\frac{c_2^2}{(4\pi)^4}W_{-1}.
\end{equation}
Then to eliminate the divergences we should shift the coupling constant $\alpha\to \alpha(\Lambda)$ and do re-summation to obtain a renormalized coupling constant $\alpha_r(\mu)$. The ansatz has the following form
\begin{equation}
\frac{1}{\alpha_r(\mu)}=\frac{1}{\alpha(\Lambda)}+\beta_1L+\beta_2\alpha(\Lambda)L+O(\alpha^2(\Lambda)),
\end{equation}
where $\beta_1$ and $\beta_2$ are the first two $\beta$-function coefficients from the Gell-Mann--Low equation 
\begin{equation}
\Lambda\frac{d\alpha(\Lambda)}{d\Lambda}=\beta(\alpha(\Lambda))=\beta_1\alpha^2(\Lambda)+\beta_2\alpha^3(\Lambda)+\ldots
\end{equation}
Thereby
\begin{equation}
\beta_1=-\frac{11}{6}\frac{c_2}{(4\pi)^2}
,\,\,\,\,\,\,
\beta_2=-\frac{9}{8}\frac{c_2^2}{(4\pi)^4}.
\end{equation}
The last $\beta_1$-function coefficient matches the well-known value \cite{13} after the following transition $\alpha(\Lambda)\to g(\Lambda)$.

\section{Green function and regularization}
\label{s4}
In this section we give a basic information on the Green function properties and applications that are necessary for calculations. This part is not included in the Appendix of the article because it contains important definitions.

\subsection{Notations}
Here we introduce several important notations that will be used throughout the rest of the work. First of all let us remind matrix analogs of the background field and the field strength
\begin{equation*} 
(B_{\mu}^{})^{ab}=f^{acb}B_{\mu}^{\,c},\,\,\,\,\,\,(F_{\mu\nu}^{})^{ab}=f^{acb}F_{\mu\nu}^{\,c}.
\end{equation*}
Then we introduce three types of derivatives: if $f\in C^{\infty}(\mathbb{R})$ is a matrix-valued function, then
\begin{equation*} 
\overrightarrow{D}_{x^{\mu}}f(x)=\partial_{x^{\mu}}f(x)+B_{\mu}(x)f(x),\,\,\,\,\,\,
f(x)\overleftarrow{D}_{x^{\mu}}=\partial_{x^{\mu}}f(x)-f(x)B_{\mu}(x),
\end{equation*}
and
\begin{equation*} 
\nabla_{x^{\mu}}f(x)=\partial_{x^{\mu}}f(x)+[B_{\mu}(x),f(x)].
\end{equation*}
It is easy to check that the following operator equality
\begin{equation}
\overrightarrow{D}_\mu\overrightarrow{D}_\nu-\overrightarrow{D}_\nu\overrightarrow{D}_\mu=F_{\mu\nu}
\end{equation}
holds true. By symbol \textquotedblleft$\mathrm{tr}$\textquotedblright\,we mean a matrix trace that is convolution of the top Lie group indices. An operator \textquotedblleft$\mathrm{Tr}$\textquotedblright\,\,is applicable to an operator (on $\mathfrak{g}$) valued smooth function of two variables $\phi\in C^{\infty}(\mathbb{R}^2)$ and it is defined by the equality
\begin{equation*} 
\mathrm{Tr}\,[\phi(x,y)]=\int_{\mathbb{R}^4}d^4y\,\bigg|_{x=y}\mathrm{tr}\,\phi(x,y).
\end{equation*}
If the function $\phi$ has a \textquotedblleft good\textquotedblright\,\,enough behaviour at infinity, then we have
\begin{equation} 
\mathrm{Tr}\,[\overrightarrow{D}_{x^{\mu}}\phi(x,y)]=-\mathrm{Tr}\,[\phi(x,y)\overleftarrow{D}_{y^{\mu}}].
\end{equation}
Also we use multi-index $(x-y)^{\mu_1\ldots\mu_k}=(x-y)^{\mu_1}\cdot\ldots\cdot(x-y)^{\mu_k}$
and the next symbols: 
\begin{equation}
\varkappa(x,y)=(x-y)^{\sigma_1\sigma_2}F_{\sigma_1\mu}^a(y)F_{\sigma_2 \mu}^a(y)=(x-y)^{\sigma_1\sigma_2}\rho_{\sigma_1\sigma_2}(y),\,\,\,\,\,\,\rho(y)=\rho_{\mu\mu}(y).
\end{equation}

\subsection{Green function expansion}
All calculations are based on using an asymptotic expansion for the Green functions (\ref{green}) when its arguments are close enough, $x\sim y$. Let us consider an abstract Laplace-type operator $\mathcal{A}$ in the four-dimensional space. An asymptotic expansion for its Green function $\mathcal{G}$ can be found by using heat kernel method (see \cite{28,29,30,31}) and has the following form
\begin{equation}
\label{eq21}
\mathcal{G}_{}(x,y)=\int_0^{+\infty}d\tau\,\left[\frac{e^{-r^2/4\tau}}{(4\pi\tau)^{2}}\bigg(\sum_{k=0}^{+\infty}\tau^k\mathfrak{a}_{k}(x,y)\bigg)-P_0(x,y)\right],
\end{equation}
where $P_0(x,y)$ is a projector on a space of zero modes \cite{29}, $\mathfrak{a}_{j}(x,y)$ are Seeley--DeWitt coefficients, and $r=|x-y|$. So the asymptotic equals to
\begin{equation}
\label{eq22} \mathcal{G}_{}(x,y)=\frac{\mathfrak{a}_{0}(x,y)}{4\pi^2r^2}-\frac{\ln(r^2\mu^2)}{16\pi^2}\mathfrak{a}_{1}(x,y)+\frac{r^2\ln(r^2\mu^2)}{64\pi^2}\mathfrak{a}_{2}(x,y)+\mathcal{PS}(x,y)+o(r^3),
\end{equation}
where $\mathcal{PS}(x,y)$ is a non-local regular part, and $\mu$ is an auxiliary dimensional parameter. Due to the equality $\mathcal{A}\,\mathcal{G}(x,y)=\mathbb{1}\delta(x-y)$, we obtain the following relation 
\begin{equation}
\label{eq1}
\left.[\mathcal{A}\,\mathcal{PS}(x,y)]\right|_{x=y}-\frac{3}{16\pi^2}\mathfrak{a}_{2}(y,y)=0.
\end{equation}
Let us concretize the formulas for our cases. Notations for Seeley--DeWitt coefficients of Green functions  $G_{1\mu\nu}$ and $G_{0}$ differ in the presence of the bottom Greek indices. We hope this does not cause confusion.

\paragraph{Green function $G_{1\mu\nu}$.} In this case the main Seeley--DeWitt coefficient is equal to the path-ordered exponential and is decomposed as $a_{0\mu\nu}(x,y)=\delta_{\mu\nu}\Phi(x,y)$, where
\begin{equation}
\label{eq6}
\Phi(x,y)=1+\sum_{k=1}^{\infty}\frac{(x-y)^{\sigma_1\ldots\sigma_k}}{k!}
\bigg(1\overleftarrow{D}_{y^{\sigma_1}}\ldots\overleftarrow{D}_{y^{\sigma_k}}\bigg),
\end{equation}
which has the following properties
\begin{equation*}
\overrightarrow{D}_{x^\mu}\Phi(x,y)=\frac{1}{2}(x-y)^\nu F_{\nu\mu}(y)+O(|x-y|^2),
\end{equation*}
\begin{equation*}
\Phi(x,y)\overleftarrow{D}_{y^\mu}=\frac{1}{2}(x-y)^\nu F_{\nu\mu}(y)+O(|x-y|^2).
\end{equation*}
Then the first and the second coefficients are calculated in the \cite{31,32,33} and have forms
\begin{align}
\nonumber
a_{1\mu\nu}(x,y)&=2F_{\mu\nu}
+(x-y)^{\sigma_1}\left(\nabla_{\sigma_1}F_{\mu\nu}+\frac{1}{6}\delta_{\mu\nu}\nabla_{\rho}F_{\sigma_1\rho}-2B_{\sigma_1}F_{\mu\nu}\right)\\\label{eq7}
&+(x-y)^{\sigma_1\sigma_2}
\left(\frac{\delta_{\mu\nu}}{12}F_{\sigma_1\rho}F_{\sigma_2\rho}+\frac{\delta_{\mu\nu}}{24}\nabla_{(\rho}\nabla_{\sigma_1)}F_{\sigma_2\rho}+\frac{1}{3}\nabla_{\sigma_1}\nabla_{\sigma_2}F_{\mu\nu}\right.\\
\nonumber
&-\left.\frac{\delta_{\mu\nu}}{6}B_{\sigma_1}\nabla_\rho F_{\sigma_2\rho}-(B_{\sigma_1}\overleftarrow{D}_{\sigma_2})F_{\mu\nu}-B_{\sigma_1}\nabla_{\sigma_2}F_{\mu\nu}\right) +O(|x-y|^3),
\end{align}
\begin{equation}
a_{2\mu\nu}(x,y)=2F_{\mu\rho}F_{\rho\nu}+\frac{\delta_{\mu\nu}}{12}F_{\sigma\rho}F_{\sigma\rho}+\frac{1}{3}\nabla_{\rho}\nabla_{\rho}F_{\mu\nu}+O(|x-y|),
\end{equation}
where all coefficients of Taylor expansions are functions of the variable $y$, and small brackets in the second line denote symmetrization without division by 2. It is easy to verify that
\begin{equation*}
\mathrm{tr}\,[a_{2\mu\mu}(y,y)]=\frac{5}{3}c_2\rho(y),
\end{equation*}
so we can rewrite the equality (\ref{eq1}) in the following form
\begin{equation}
\label{eq2}
\mathrm{tr}\left.[\overrightarrow{D}_{x^\mu}\overrightarrow{D}_{x^\mu} PS_{\nu\nu}(x,y)]\right|_{x=y}=2\,\mathrm{tr}\,[F_{\mu\nu}(y)PS_{\mu\nu}(y,y)]-\frac{5c_2\rho(y)}{2^4\pi^2}.
\end{equation}

\paragraph{Green function $G_{0}$.} For this case all previous steps are also possible, so we only declare the results:
\begin{equation}
a_0(x,y)=\Phi(x,y),\,\,\,\,\,\,
a_1(x,y)=\frac{1}{4}a_{1\mu\mu}(x,y),\,\,\,\,\,\,
a_2(x,y)=\frac{1}{12}F_{\sigma\rho}(y)F_{\sigma\rho}(y)+O(|x-y|),
\end{equation}
\begin{equation}
\mathrm{tr}\,[a_{2}(y,y)]=-\frac{1}{12}c_2\rho(y),\,\,\,\,\,\,
\mathrm{tr}\left.[\overrightarrow{D}_{x^\mu}\overrightarrow{D}_{x^\mu}PS(x,y)]\right|_{x=y}=\frac{c_2\rho(y)}{2^6\pi^2}.
\end{equation}
It should be also noted that the relation holds true
\begin{equation}
a_{1\nu\beta}(x,y)+a_{1\beta\nu}(x,y)=\frac{1}{2}\delta_{\nu\beta}a_{1\rho\rho}(x,y).
\end{equation}

\paragraph{Additional properties.} Let us note one useful relation for the regular part $PS_{\mu\nu}$ of the function $G_{1\mu\nu}$. First of all we remind that the following expressions
\begin{equation}
\label{eq3}
\mathrm{tr}\left.[\overrightarrow{D}_{x^\mu}\overrightarrow{D}_{x^\mu}PS_{\nu\nu}(x,y)]\right|_{x=y}+2\,\mathrm{tr}\,[F_{\mu\nu}(y)PS_{\nu\mu}(y,y)],\,\,\,\,\,\,
\mathrm{tr}\left.[\overrightarrow{D}_{x^\mu}\overrightarrow{D}_{x^\nu}PS_{\nu\mu}(x,y)]\right|_{x=y}
\end{equation}
are invariant with respect to the transition to the Fock--Schwinger gauge condition \cite{33}. In particular this means that the covariant derivative equals to
\begin{equation}
\label{eq50}
\overrightarrow{D}_{x^\mu}=\partial_{x^\mu}+\frac{1}{2}(x-y)^{\nu}F_{\nu\mu}(y)+O(|x-y|^2).
\end{equation}
Then we need to recall that Taylor coefficients of $PS_{\mu\nu}(x,y)$ are local polynomials. At the same time it should be equal to zero when the background field vanishes. It means that we can take the ansatz in the form (see Appendix C)
\begin{align}
\nonumber
PS_{\mu\nu}(x,y)&=aF_{\mu\nu}(y)+(x-y)^{\rho}b_{\rho\mu\nu}(y)\\\label{eq52}
&+cr^2\delta_{\mu\nu}F_{\sigma\rho}F_{\sigma\rho}+dr^2F_{\mu\rho}F_{\rho\nu}+e\delta_{\mu\nu}(x-y)^{\sigma\rho}F_{\beta\sigma}F_{\beta\rho}\\\nonumber
&+(x-y)^{\sigma\rho}C_{\sigma\rho\mu\nu}(y)+o(|x-y|^2),
\end{align}
where a function $b_{\rho\mu\nu}(y)$ is not interesting, and $C_{\sigma\rho\mu\nu}(y)$ is traceless. We should note that coefficients like $(x-y)^{\mu\nu}F_{\sigma\rho}F_{\sigma\rho}$, $(x-y)^{\mu\sigma}F_{\sigma\rho}F_{\rho\nu}$, or $(x-y)^{\nu\sigma}F_{\sigma\rho}F_{\rho\mu}$ can not appear by construction (see also \cite{31} and \cite{33}). Then we rewrite the expressions from (\ref{eq3}) as
\begin{equation*}
\mathrm{tr}\left.[\partial_{x^\mu}\partial_{x^\mu}PS_{\nu\nu}(x,y)\right|_{x=y}+2\,\mathrm{tr}\,[F_{\mu\nu}(y)PS_{\nu\mu}(y,y)]=c_2\rho(y)(2a-8e-8d-32c),
\end{equation*}
and
\begin{equation*}
\mathrm{tr}\left.[\partial_{x^\mu}\partial_{x^\nu}PS_{\nu\mu}(x,y)\right|_{x=y}+\frac{1}{2}\mathrm{tr}\,[F_{\mu\nu}(y)PS_{\nu\mu}(y,y)]=\frac{c_2\rho(y)}{4}(2a-8e-8d-32c).
\end{equation*}
Therefore, by using the relation (\ref{eq2}), we obtain the following equality
\begin{equation}
\label{eq47}
\mathrm{tr}\left.[\overrightarrow{D}_{x^\mu}\overrightarrow{D}_{x^\nu}PS_{\nu\mu}(x,y)]\right|_{x=y}=-\frac{5c_2\rho(y)}{2^6\pi^2}.
\end{equation}
Also we can note that from the formula (\ref{eq52}) the identity follows
\begin{equation}
\label{eq48}
\mathrm{tr}\,[PS_{\nu\nu}(y,y)]=0.
\end{equation}

\subsection{Cutoff momentum}
\label{cutoff}
\paragraph{Regularization.}
In the present paper we use a special type of cutoff regularization. However, we do not make the transition to a momentum representation. We deform our Green functions by introducing a parameter $\Lambda$ that has dimension of momentum. The rules are following:
\begin{equation*}
r\to r_{\Lambda}=
\begin{cases}
r &,\,1/\Lambda\leqslant r;\\
1/\Lambda &,\,0\leqslant r<1/\Lambda.
\end{cases}
\end{equation*}
It is obvious that $r_{\Lambda}\to r$ when $\Lambda\to+\infty$. After substitution $r_{\Lambda}$ instead of $r$ in the formula (\ref{eq21}) the Green function has the following expansion
\begin{equation}
\label{eq54} \mathcal{G}_{\Lambda}(x,y)=\frac{\mathfrak{a}_{0}(x,y)}{4\pi^2r_{\Lambda}^2}-\frac{\ln(r_{\Lambda}^2\mu^2)}{16\pi^2}\mathfrak{a}_{1}(x,y)+\frac{r_{\Lambda}^2\ln(r_{\Lambda}^2\mu^2)}{64\pi^2}\mathfrak{a}_{2}(x,y)+\mathcal{PS}_{\Lambda}(x,y)+o(r_{\Lambda}^3),
\end{equation}
where the Seeley--DeWitt coefficients are not deformed. This is required in order not to change important functional properties. In particular we have $\mathcal{PS}_{\Lambda}(x,y)=\mathcal{PS}(x,y)$ for $r\geqslant 1/\Lambda$. At the same time we have a convergence $\mathcal{G}_{\Lambda}(x,y)\to\mathcal{G}_{}(x,y)$, when $\Lambda\to+\infty$, in the sense of generalized functions.

\paragraph{Singularities.}
In the work we consider only infrared singularities. It means that $x\sim y$. We label an equal sign \textquotedblleft$=$\textquotedblright\,\,by an index \textquotedblleft$IR\,$\textquotedblright\,\,if the right and left parts have the same singularities.
To analyze integrals, it is enough to know the following three relations:
\begin{equation*}
\int^{+\infty}_{1/\Lambda}\frac{dr}{r^3}\stackrel{IR}{=}\frac{1}{2}\Lambda^2,\,\,\,\,\,\,
\int^{1/\mu}_{1/\Lambda}\frac{dr}{r}\stackrel{IR}{=}\ln(\Lambda/\mu)=L,\,\,\,\,\,\,
\int^{1/\mu}_{1/\Lambda}\frac{dr\,\ln(r\mu)}{r}\stackrel{IR}{=}-\frac{1}{2}L^2,
\end{equation*}
where the dimensional parameter $\mu$ can be selected the same as in the formula (\ref{eq54}), since this does not affect the singularity values in our calculations.
Thus we can consider integrals of this type
\begin{equation}
\label{eq39}
\int_{\mathbb{R}^4}d^4x\,\frac{(x-y)^{\mu\nu}}{r^6}\stackrel{IR}{=}\frac{\delta^{\mu\nu}S^3L}{4},\,\,\,\,\,\,
\int_{\mathbb{R}^4}d^4x\,\frac{(x-y)^{\mu\nu}\ln(r\mu)}{r^6}\stackrel{IR}{=}-\frac{\delta^{\mu\nu}S^3L^2}{8},
\end{equation}
where $S^3=2\pi^2$ is the surface area of a sphere in $\mathbb{R}^4$ with the unit radius.
It should be noted that if an integrand contains $(x-y)^{\mu_1\ldots\mu_k}$ with odd degree, then the integral is equal to zero due to the symmetry.

\section{The first correction}
\label{s5}
One-loop correction is described by the second term on the right hand side of (\ref{eq20}). Let us use the regularization and find a singular part. Thus we remind that after regularization $\ln\det(\mathcal{A})$ has the form
\begin{equation}
\label{eq55}
-\mathrm{Tr}\int_0^{+\infty}\frac{d\tau}{\tau}\,(4\pi\tau)^{-2}e^{-r_{\Lambda}^2/4\tau}\bigg(\sum_{k=0}^{+\infty}\tau^k\mathfrak{a}_{k}(x,y)\bigg).
\end{equation}
Now we note that an asymptotic for the integral can be calculated in the same way as in the (\ref{eq22}), but by using formula for the six-dimensional case multiplied by $4\pi$. So the expression under the trace has the following singular part
\begin{equation*}
-\frac{\mathfrak{a}_{0}(x,y)}{\pi^2r_{\Lambda}^4}-\frac{\mathfrak{a}_{1}(x,y)}{4\pi^2r_{\Lambda}^2}+\frac{\ln(r_{\Lambda}^2\mu^2)}{16\pi^2}\mathfrak{a}_{2}(x,y).
\end{equation*}
Then, due to the last asymptotic and the equality $\mathrm{tr}\,[a_{1\nu\nu}(y,y)]=0$, we can rewrite the one-loop singular part from (\ref{eq20}) as
\begin{equation*}
\int_{\mathbb{R}^4}d^4y \left[\frac{1}{2}\frac{\mathrm{tr}\,[a_{2\mu\mu}(y,y)]\ln(\frac{\mu}{\Lambda})^2}{16\pi^2}-\frac{\mathrm{tr}\,[a_2(y,y)]\ln(\frac{\mu}{\Lambda})^2}{16\pi^2}\right],
\end{equation*}
where $\mathrm{tr}\,[a_{2\mu\mu}(y,y)]=5c_2\rho(y)/3$ and $\mathrm{tr}\,[a_2(y,y)]=-c_2\rho(y)/12$. Finally, if $L=\ln(\Lambda/\mu)$ one can get
\begin{equation}
-\frac{11L}{6}\frac{c_2}{(4\pi)^2}W_{-1}.
\end{equation}

\section{The second correction}
\label{s6}
 Two-loop calculation is time-consuming enough procedure. So it is convenient to introduce a special case formalism \cite{35} that allows us to write formulas in a very compact way. Let $A$, $B$, and $C$ are smooth Lie group valued functions in the matrix representation. So we can define
\begin{equation}
\label{eq25}
(A,B,C)=f^{ace}A^{ab}B^{cd}C^{eg}f^{bdg}.
\end{equation}
Let us note some properties:
\begin{equation}
(A,B,C)=(B,A,C)=(A,C,B),
\end{equation}
and, if $X^{ab}=f^{acb}x^c$, then, due to the formulas from (\ref{eq4}), we have
\begin{equation}
\label{eq8}
(XA,B,C)=-(A,XB,C)-(A,B,XC).
\end{equation}
Now we are ready to consider all  possible two-loop corrections and reduce them by using symmetry properties.

\paragraph{$\mathbf{\Gamma}^2_3$ term:}
Firstly, we consider only $\mathrm{1PI}$ contribution from $\frac{g^2}{2}\Gamma_3^2$ to the second line of the formula (\ref{eq20}). It is constructed by the expression
\begin{multline*}
\label{Gamma_3}
\frac{g^2}{2}\left[
\int_{\mathbb{R}^4}d^4x\,f^{abc}
\frac{\delta}{\delta J^a_\alpha(x)}\frac{\delta}{\delta J^b_\beta(x)}D^{cd}_\alpha\frac{\delta}{\delta J^d_\beta(x)}
\right]^2\\
\frac{1}{2^33!}\left(
\int_{\mathbb{R}^4}d^4y_1\int_{\mathbb{R}^4}d^4y_2\, \mathrm{tr}\,[J_\mu(y_1)G_{1\mu\nu}(y_1,y_2)J_\nu(y_2)]\right)^3.
\end{multline*}
Diagrammatically it is related to the first graph on the Figure \ref{loop}. We can permute the Green functions, so such kind of symmetry gives $3!$ identical terms. Also the Green function has the following symmetry
\begin{equation} 
\label{eq35}
G_{1\mu\nu}^{\,\,\,ab}(x,y)=G_{1\nu\mu}^{\,\,\,ba}(y,x), 
\end{equation}
therefore we have the factor $2^3$.
Thus we should find only six different types.
Let $G_{1\mu\nu}=G_{1\mu\nu}(x,y)$, the right derivative $\overrightarrow{D}_\alpha$ acts by the first argument, and the left derivative $\overleftarrow{D}_\alpha$ acts by the second argument.
So we have the following terms
\begin{equation}
\label{eq26}
\,\,\,\,\,\,\,
(G_{1\mu\alpha},G_{1\nu\beta},\overrightarrow{D}_\nu G_{1\mu\alpha}\overleftarrow{D}_\beta)
\end{equation}
\begin{equation}
\label{eq27}
-(G_{1\mu\alpha},G_{1\nu\beta}\overleftarrow{D}_\alpha ,\overrightarrow{D}_\mu G_{1\nu\beta})
\end{equation}
\begin{equation}
\label{eq28}
-(G_{1\mu\beta},G_{1\nu\alpha},\overrightarrow{D}_\mu G_{1\nu\beta}\overleftarrow{D}_\alpha)
\end{equation}
\begin{equation}
\label{eq29}
+(G_{1\mu\beta}, G_{1\nu\beta}\overleftarrow{D}_\alpha,\overrightarrow{D}_\mu G_{1\nu\alpha})
\end{equation}
\begin{equation}
\label{eq30}
+( G_{1\mu\beta} \overleftarrow{D}_\alpha,G_{1\nu\alpha},\overrightarrow{D}_\mu G_{1\nu\beta})
\end{equation}
\begin{equation}
\label{eq31}
-( G_{1\mu\beta}\overleftarrow{D}_\alpha,G_{1\nu\beta},\overrightarrow{D}_\mu G_{1\nu\alpha})
\end{equation}
under the operator $\frac{g^2}{2}\int_{\mathbb{R}^4}d^4x\int_{\mathbb{R}^4}d^4y$. But the last expression contains the second derivative of the Green function. To eliminate them let us use the formula (\ref{eq8}) in the forms
\begin{multline}
\label{eq32}
\int_{\mathbb{R}^4}d^4x\int_{\mathbb{R}^4}d^4y\,(G_{1\mu\alpha},G_{1\nu\beta},\overrightarrow{D}_\nu G_{1\mu\alpha}\overleftarrow{D}_\beta)\\=
-\int_{\mathbb{R}^4}d^4x\int_{\mathbb{R}^4}d^4y\,( G_{1\mu\alpha},\overrightarrow{D}_\nu G_{1\nu\beta},G_{1\mu\alpha}\overleftarrow{D}_\beta)\\
-\int_{\mathbb{R}^4}d^4x\int_{\mathbb{R}^4}d^4y\,(\overrightarrow{D}_\nu G_{1\mu\alpha},G_{1\nu\beta},G_{1\mu\alpha}\overleftarrow{D}_\beta),
\end{multline}
\begin{multline}
\label{eq33}
-\int_{\mathbb{R}^4}d^4x\int_{\mathbb{R}^4}d^4y\,(G_{1\mu\beta},G_{1\nu\alpha},\overrightarrow{D}_\mu G_{1\nu\beta}\overleftarrow{D}_\alpha)\\
-\int_{\mathbb{R}^4}d^4x\int_{\mathbb{R}^4}d^4y\,(G_{1\mu\beta}\overleftarrow{D}_\alpha,G_{1\nu\beta},\overrightarrow{D}_\mu G_{1\nu\alpha})\\
=2\int_{\mathbb{R}^4}d^4x\int_{\mathbb{R}^4}d^4y\,(G_{1\mu\beta},G_{1\nu\beta}\overleftarrow{D}_\alpha,\overrightarrow{D}_\mu G_{1\nu\alpha})\\
-\int_{\mathbb{R}^4}d^4x\int_{\mathbb{R}^4}d^4y\,(G_{1\mu\alpha},\overrightarrow{D}_\nu G_{1\nu\alpha}, G_{1\mu\beta}\overleftarrow{D}_\beta).
\end{multline}
Applying the equality (\ref{eq32}) to the term (\ref{eq26}) and the relation (\ref{eq33}) to the expressions (\ref{eq28}) and (\ref{eq31}), we get only the following four contributions instead of (\ref{eq26})-(\ref{eq31})
\begin{equation}
\label{eq41}
\mathcal{J}_1=2g^2\int_{\mathbb{R}^4}d^4x\int_{\mathbb{R}^4}d^4y\,
(G_{1\mu\alpha},G_{1\nu\beta}\overleftarrow{D}_\alpha,\overrightarrow{D}_\nu G_{1\mu\beta}),
\end{equation}
\begin{equation}
\label{eq42}
\mathcal{J}_2=-g^2\int_{\mathbb{R}^4}d^4x\int_{\mathbb{R}^4}d^4y\,
(G_{1\mu\alpha},G_{1\nu\beta}\overleftarrow{D}_\alpha ,\overrightarrow{D}_\mu G_{1\nu\beta}),
\end{equation}
\begin{equation}
\label{eq43}
\mathcal{J}_3=-\frac{g^2}{2}\int_{\mathbb{R}^4}d^4x\int_{\mathbb{R}^4}d^4y\,
(G_{1\mu\alpha},\overrightarrow{D}_\nu G_{1\nu\beta},G_{1\mu\alpha}\overleftarrow{D}_\beta),
\end{equation}
\begin{equation}
\label{eq44}
\mathcal{J}_4=-\frac{g^2}{2}\int_{\mathbb{R}^4}d^4x\int_{\mathbb{R}^4}d^4y\,
( G_{1\mu\alpha},\overrightarrow{D}_\nu G_{1\nu\alpha},G_{1\mu\beta}\overleftarrow{D}_\beta).
\end{equation}

\paragraph{$\mathbf{\Omega}^2$ term:}  In the same way let us find a $\mathrm{1PI}$ ghost contribution from
 \begin{multline*}
\frac{g^2}{2}\left[
\int_{\mathbb{R}^4}d^4x\,f^{abc}
\frac{\delta}{\delta J^a_\alpha(x)}\frac{\delta}{\delta \bar{b}^{\,b}(x)}D^{cd}_\alpha \frac{\delta}{\delta b^{\,d}(x)}
\right]^2\\
\frac{1}{2}
\left(\int_{\mathbb{R}^4}d^4y_1\int_{\mathbb{R}^4}d^4y_2\, \mathrm{tr}\,[J_\mu(y_1)G_{1\mu\nu}(y_1,y_2)J_\nu(y_2)]\right)\\
\frac{1}{2}
\left(\int_{\mathbb{R}^4}d^4x_1\int_{\mathbb{R}^4}d^4x_2\,\mathrm{tr}\,[b(x_1)G_0(x_1,x_2)\bar{b}(x_2)]\right)^2.
\end{multline*}
In this case the symmetry factors are $2$, due to the second order of  $G_0$, and $2$ because of the symmetry (\ref{eq35}).  Therefore we have only one diagram. Using the notations from the previous section and $G_0^{*ab}(x,y)=G_0^{ba}(y,x)$ we get
\begin{equation}
\label{eq45}
\mathcal{J}_5=\frac{g^2}{2}\int_{\mathbb{R}^4}d^4x\int_{\mathbb{R}^4}d^4y\,
(G_{1\mu\alpha},G_0^*\overleftarrow{D}_\alpha,\overrightarrow{D}_\mu G_0).
\end{equation}

\paragraph{$\mathbf{\Gamma}_4$ term:} The last contributions follow from
\begin{multline*}
-\frac{g^2}{4}\left[\int_{\mathbb{R}^4}d^4x\,f^{cab}f^{cde}\frac{\delta}{\delta J^a_\mu(x)}\frac{\delta}{\delta J^b_\nu(x)}\frac{\delta}{\delta J^d_\mu(x)}\frac{\delta}{\delta J^e_\nu(x)}\right]\\
\frac{1}{2^3}
\left(\int_{\mathbb{R}^4}d^4y_1\int_{\mathbb{R}^4}d^4y_2\, \mathrm{tr}\,[J_\mu(y_1)G_{1\mu\nu}(y_1,y_2)J_\nu(y_2)]\right)^2.
\end{multline*}
Because of the permutations of $G_{1\mu\nu}$ and its symmetry (\ref{eq35}), we have $2^3$ similar terms. So only three different cases are possible
\begin{multline}
\label{eq40}
\mathcal{J}_6=-\frac{g^2}{4}\int_{\mathbb{R}^4}d^4y\bigg(f^{cab}f^{cde}G_{1\mu\nu}^{\,\,\,ab}(y,y)G_{1\mu\nu}^{\,\,\,de}(y,y)\\
+f^{cab}f^{cde}G^{\,\,\,ae}_{1\mu\nu}(y,y)G^{\,\,\,db}_{1\mu\nu}(y,y)+f^{cab}f^{cde}G_{1\mu\mu}^{\,\,\,ad}(y,y)G_{1\nu\nu}^{\,\,\,be}(y,y)\bigg).
\end{multline}

\paragraph{Auxiliary notation.}
We notice that expressions $\mathcal{J}_1,\ldots,\mathcal{J}_5$ contain a product of three Green functions. So it is convenient to use a special type of notation.
Let us number the summands in the Green function decomposition (\ref{eq22}) from left to right. Then by symbol
\begin{equation}
\label{eq36}
I^{n}_{i,j,k}\,,\,\,\,\,\,\,n\in\{1,\ldots,5\},\,\,i,j,k\in\{1,\ldots,4\}
\end{equation}
we denote a contribution to the $\mathcal{J}_n$ such that the left Green function is replaced by the i-th item from (\ref{eq22}), and the middle and the right ones are substituted by j-th and k-th items respectively. For example,
\begin{equation*}
I^1_{1,3,1}=
2g^2\int_{\mathbb{R}^4}d^4x\int_{\mathbb{R}^4}d^4y
\left(\frac{\delta_{\mu\alpha}\Phi}{4\pi^2r^2},\left[\frac{r^2\ln(r\mu)}{32\pi^2}a_{2\nu\beta}\right]\overleftarrow{D}_\alpha,\overrightarrow{D}_\nu\frac{\delta_{\mu\beta}\Phi}{4\pi^2r^2}\right).
\end{equation*}
In particular it means that we can calculate some contributions separately, see Appendix A, and then use them. Further for simplicity let us also denote $\Phi=\Phi(x,y)$, $a_{1\mu\nu}=a_{1\mu\nu}(x,y)$, $a_{2\mu\nu}=a_{2\mu\nu}(x,y)$, $F_{\mu\nu}=F_{\mu\nu}(y)$, $\rho_{\mu\rho}=\rho_{\mu\rho}(y)$, $\varkappa=\varkappa(x,y)$, $PS_{\mu\nu}=PS_{\mu\nu}(x,y)$, and $PS=PS(x,y)$.

\subsection{Calculation of $\mathcal{J}_1$}
In this section we give calculation of the first $\mathcal{J}$-component in detail. Other components are described in Section \ref{oth} and Appendix B.
As it was noted above, we are planning to split
\begin{equation}
\mathcal{J}_1=2g^2\int_{\mathbb{R}^4}d^4x\int_{\mathbb{R}^4}d^4y\,
\big(G_{1\mu\alpha},G_{1\nu\beta}\overleftarrow{D}_\alpha,\overrightarrow{D}_\nu G_{1\mu\beta}\big)
\end{equation}
into several parts. Due to the notation (\ref{eq35}) we should calculate $I^{n}_{i,j,k}$ only for several combinations that give singularities. We can distribute them into two groups. The first one components are constructed only by Seeley--DeWitt coefficients
\begin{equation*}
\{1,1,2\},\,\{1,2,1\},\,\{2,1,1\},\,\{1,1,3\},\,\{1,3,1\},\,\{3,1,1\},\,\{1,2,2\},\,\{2,1,2\},\,\{2,2,1\},\,\{1,1,1\}.
\end{equation*}
They are named \textbf{\textit{local}} expressions. Contributions to the second group 
\begin{equation*}
\{1,1,4\},\,\{1,4,1\},\,\{4,1,1\},\,\{4,2,1\},\,\{4,1,2\}
\end{equation*}
contain regular part, $PS_{\mu\nu}$ or $PS$, of the Green function
and are called \textbf{\textit{non-local}} elements.
\paragraph{Local contributions.}
Let us show how the procedure of calculation works. We take $I^1_{1,1,2}$ and by using formulas from Appendix A expand the integrand in a Taylor series: if $0\leqslant r<1/\Lambda$, then
\begin{equation}
\label{eq38}
\left(\frac{\delta_{\mu\alpha}\Phi}{4\pi^2r^2},\frac{\delta_{\nu\beta}\Phi}{4\pi^2r^2}\overleftarrow{D}_\alpha,\overrightarrow{D}_\nu\left[-\frac{\ln(r\mu)a_{1\mu\beta}}{8\pi^2}\right]\right)=
\Lambda^4\ln(\Lambda/\mu)O(r),
\end{equation}
and if $r\geqslant1/\Lambda$, then
\begin{align}
\label{eq37}
\left(\frac{\delta_{\mu\alpha}\Phi}{4\pi^2r^2},\frac{\delta_{\nu\beta}\Phi}{4\pi^2r^2}\overleftarrow{D}_\alpha,\overrightarrow{D}_\nu\left[-\frac{\ln(r\mu)a_{1\mu\beta}}{8\pi^2}\right]\right)=&
-\frac{1}{2^7\pi^6 r^6}(\Phi,\Phi\overleftarrow{D}_\alpha,(x-y)^\beta a_{1\alpha\beta})\\\nonumber
&-\frac{1}{2^6\pi^6 r^8}(\Phi,\Phi,(x-y)^{\alpha\beta}a_{1\alpha\beta})\\\nonumber
&-\frac{\ln(r\mu)}{2^6 \pi^6 r^6}(\Phi,\Phi,(x-y)^\alpha\overrightarrow{D}_\beta a_{1\alpha\beta})+\ldots\\\nonumber
&=\frac{c_2^2\varkappa}{2^7\pi^6r^6}\left(\frac{2}{3}+\ln(r\mu)\frac{7}{3}\right)+\ldots,
\end{align}
where the dots represent an unimportant part.
Now we should note that the term (\ref{eq38}) actually does not give an infrared contribution because of the relation $5\int_0^{1/\Lambda}dr\,r^4=\Lambda^{-5}$. After applying the formula (\ref{eq39}), the term (\ref{eq37}) gives nonzero contribution
\begin{equation}
I^1_{1,1,2}\stackrel{IR}{=}\frac{g^2c_2^2S^3W_{-1}}{2^7\pi^6}
\left(\frac{L}{3}-\frac{7L^2}{12}\right).
\end{equation}
Repeating step by step we can find contributions from other components:
\begin{equation}
I^1_{1,2,1}\stackrel{IR}{=}\frac{g^2c_2^2S^3W_{-1}}{2^7\pi^6}
\left(-\frac{L}{6}+\frac{5L^2}{12}\right),\,\,\,\,\,\,
I^1_{2,1,1}\stackrel{IR}{=}\frac{g^2c_2^2S^3W_{-1}}{2^7\pi^6}
\left(-\frac{5L^2}{12}\right),
\end{equation}
\begin{equation}
I^1_{1,1,3}+I^1_{1,3,1}+I^1_{3,1,1}\stackrel{IR}{=}\frac{g^2c_2^2S^3W_{-1}}{2^7\pi^6}
\left(\frac{5L}{6}-\frac{5L^2}{12}\right),
\end{equation}
\begin{equation}
I^1_{1,2,2}+I^1_{2,1,2}+I^1_{2,2,1}\stackrel{IR}{=}\frac{g^2c_2^2S^3W_{-1}}{2^7\pi^6}
\left(-\frac{L}{2}\right).
\end{equation}
The term $I^1_{1,1,1}$ is a little bit different and has decompositions:
if $0\leqslant r<1/\Lambda$, then
\begin{equation}
\left(\frac{\delta_{\mu\alpha}\Phi}{4\pi^2r^2},\frac{\delta_{\nu\beta}\Phi}{4\pi^2r^2}\overleftarrow{D}_\alpha, \overrightarrow{D}_\nu\frac{\delta_{\mu\beta}\Phi}{4\pi^2r^2}\right)=
\Lambda^6O(r^2),
\end{equation}
and if $r\geqslant1/\Lambda$, then
\begin{align*}
\left(\frac{\delta_{\mu\alpha}\Phi}{4\pi^2r^2},\frac{\delta_{\nu\beta}\Phi}{4\pi^2r^2}\overleftarrow{D}_\alpha, \overrightarrow{D}_\nu\frac{\delta_{\mu\beta}\Phi}{4\pi^2r^2}\right)
=-\frac{c_2\dim\mathfrak{g}}{2^4\pi^6r^8}+\frac{c_2^2\varkappa}{2^7\pi^6r^6}\frac{1}{4}+\ldots,
\end{align*}
where the relations (\ref{eq9}) and (\ref{eq56}) were used. Let us denote infinite large constant $\int_{\mathbb{R}^4}d^4y=\theta$, so we have
\begin{equation}
I^1_{1,1,1}=-\Lambda^4\frac{g^2c_2S^3\theta\dim\mathfrak{g}}{2^5\pi^6}+\frac{g^2c_2^2S^3W_{-1}}{2^7\pi^6}\left(\frac{L}{8}\right).
\end{equation}
Here we should note that the first term in the last formula does not play a crucial role because it does not contain the background field. So it is cancelled due to the formula (\ref{eq46}).

\paragraph{Non-local contributions.} Let us consider the case $I^1_{1,1,4}$. It is obvious that non-zero contribution gives only $r\geqslant1/\Lambda$ part for which the expansion is
\begin{align*}
\left(\frac{\delta_{\mu\alpha}\Phi}{4\pi^2r^2},\frac{\delta_{\nu\beta}\Phi}{4\pi^2r^2}\overleftarrow{D}_\alpha,\overrightarrow{D}_\nu PS_{\mu\beta}\right)
&=r^{-6}(x-y)^{\mu}b_{\mu}^1(y)\\
&+\frac{1}{2^3\pi^4r^6}(1,1,(x-y)^{\alpha\sigma}\partial_\sigma \overrightarrow{D}_\beta PS_{\alpha\beta})\\
&-\frac{1}{2^2\pi^4r^6}(1,(x-y)^\sigma B_\sigma,(x-y)^\alpha\overrightarrow{D}_\beta PS_{\alpha\beta})+\ldots,
\end{align*}
where $b_\mu^1(y)$ is some function of $y$. So after the integration we obtain
\begin{equation}
I^1_{1,1,4}\stackrel{IR}{=}\frac{g^2c_2S^3L}{2^4\pi^4}\mathrm{Tr}\,[\overrightarrow{D}_{x^\alpha}\overrightarrow{D}_{x^\beta}PS_{\alpha\beta}(x,y)].
\end{equation}
In the analogous way we have
\begin{equation}
I^1_{1,4,1}\stackrel{IR}{=}\frac{g^2c_2S^3L}{2^4\pi^4}\mathrm{Tr}\,[\overrightarrow{D}_{x^\beta}\overrightarrow{D}_{x^\alpha}PS_{\alpha\beta}(x,y)],\,\,\,\,\,\,
I^1_{4,2,1}+I^1_{4,1,2}\stackrel{IR}{=}0.
\end{equation}
The case $\{4,1,1\}$ requires individual consideration because it has more complex decomposition for $r\geqslant1/\Lambda$:
\begin{align*}
\left(PS_{\mu\alpha},\frac{\delta_{\nu\beta}\Phi}{4\pi^2r^2}\overleftarrow{D}_\alpha,\overrightarrow{D}_\nu\frac{\delta_{\mu\beta}\Phi}{4\pi^2r^2}\right)
&=-\frac{(x-y)^{\alpha\beta}}{2^2\pi^4r^8}(PS_{\alpha\beta}(y,y),1,1)\\
&+r^{-8}(x-y)^{\mu\nu\alpha}b_{\mu\nu\alpha}^2(y)\\
&-\frac{c_2(x-y)^{\alpha\beta\rho\sigma}}{2^3\pi^4r^8}\mathrm{tr}\left.[\overrightarrow{D}_{x^\rho}\overrightarrow{D}_{x^\sigma}PS_{\alpha\beta}(x,y)]\right|_{x=y}\\
&+\frac{c_2}{2^5\pi^4r^6}(x-y)^{\alpha\sigma}\mathrm{tr}\,[PS_{\alpha\beta}(y,y)F_{\sigma\beta}(y)]\\
&+\frac{c_2}{2^5\pi^4r^6}(x-y)^{\alpha\sigma}\mathrm{tr}\,[PS_{\beta\alpha}(y,y)F_{\beta\sigma}(y)]+\ldots,
\end{align*}
where $b_{\mu\nu\alpha}^2(y)$ is also some function of $y$. After the integration we get
\begin{align}
\label{eq49}
I^1_{4,1,1}\stackrel{IR}{=}
	&-\frac{g^2c_2S^3\Lambda^2}{2^4\pi^4}\mathrm{Tr}\,[PS_{\alpha\alpha}(x,y)]
	+\frac{g^2c_2S^3L}{2^5\pi^4}\mathrm{Tr}\,[PS_{\alpha\beta}(x,y)F_{\alpha\beta}(y)]\\\nonumber
	&-\frac{g^2c_2S^3L}{3\pi^42^5}\mathrm{Tr}\,[\overrightarrow{D}_{x^\alpha}\overrightarrow{D}_{x^\alpha}PS_{\beta\beta}(x,y)]
	-\frac{g^2c_2S^3L}{3\pi^42^5}\mathrm{Tr}\,[\overrightarrow{D}_{x^\alpha}\overrightarrow{D}_{x^\beta}PS_{\alpha\beta}(x,y)]\\\nonumber
	&-\frac{g^2c_2S^3L}{3\pi^42^5}\mathrm{Tr}\,[\overrightarrow{D}_{x^\alpha}\overrightarrow{D}_{x^\beta}PS_{\beta\alpha}(x,y)],
\end{align}
where we used the integration over the unit sphere $\mathbb{S}^3$ centered at the origin ($d\sigma$ is the measure)
\begin{equation*}
\int_{\mathbb{S}^3}d\sigma\,x_ix_jx_kx_l=\frac{\pi^2}{12}(\delta_{ij}\delta_{kl}+
\delta_{ik}\delta_{jl}+\delta_{il}\delta_{jk}),\,\,\,\,\,\,
\int_{\mathbb{S}^3}d\sigma=S^3.
\end{equation*}
Then to transform non-local contributions into local we use formulas (\ref{eq2}), (\ref{eq47}), and (\ref{eq48}). For example, the term (\ref{eq49}) is equal to
\begin{equation}
I^1_{4,1,1}\stackrel{IR}{=}\frac{g^2c_2^2S^3W_{-1}}{2^7\pi^6}\left(\frac{5L}{8}\right).
\end{equation}
\paragraph{Summation of the $\mathcal{J}_1$-contributions.}
So after transformations of non-local terms into local ones and summation we obtain:
\begin{equation}
\mathcal{J}_1^{loc}-\mathcal{J}_1^{loc}\big|_{B=0}
\stackrel{IR}{=}\frac{g^2c_2^2S^3W_{-1}}{2^7\pi^6}\left(\frac{5L}{8}-L^2\right),
\end{equation}
\begin{equation}
\mathcal{J}_1^{non-loc}\stackrel{IR}{=}
\frac{g^2c_2^2S^3W_{-1}}{2^7\pi^6}\left(-\frac{5L}{8}\right)+
\frac{g^2c_2S^3L}{\pi^42^4}
\mathrm{Tr}\,[PS_{\alpha\beta}(y,y)F_{\alpha\beta}(y)],
\end{equation}
\begin{equation}
\mathcal{J}_1-\mathcal{J}_1\big|_{B=0}\stackrel{IR}{=}
-\frac{g^2c_2^2S^3W_{-1}}{2^7\pi^6}L^2+
\frac{g^2c_2S^3L}{\pi^42^4}
\mathrm{Tr}\,[PS_{\alpha\beta}(y,y)F_{\alpha\beta}(y)].
\end{equation}

\subsection{Calculation of $\mathcal{J}_6$}
The last component has a different structure.
From the formula (\ref{eq40}) it follows that $\mathcal{J}_6$ is equal to sum of three terms $I^6_{1}+I^6_{2}+I^6_{3}$. So we can calculate them separately. Let us consider the first one
\begin{equation}
I^6_{1}=-\frac{g^2}{4}\int_{\mathbb{R}^4}d^4y\,f^{cab}f^{cde}G^{\,\,\,ab}_{1\mu\nu}(y,y)G^{\,\,\,de}_{1\mu\nu}(y,y).
\end{equation}
Then using the equalities $f^{cab}\delta^{ab}=0$ and $a_{1\mu\nu}(y,y)\delta_{\mu\nu}=0$, we obtain
\begin{equation}
I^6_{1}\stackrel{IR}{=}-\frac{g^2c_2^2L^2W_{-1}}{2^6\pi^4}+\frac{g^2c_2L}{2^3\pi^2}\mathrm{Tr}\,[F_{\nu\mu}(y)PS_{\nu\mu}(y,y)].
\end{equation}
In the analogous way we have
\begin{align*}
I^6_{2}&=-\frac{g^2}{4}\int_{\mathbb{R}^4}d^4y\,f^{cab}f^{cde}G^{\,\,\,ae}_{1\mu\nu}(y,y)G^{\,\,\,db}_{1\mu\nu}(y,y)\\
&\stackrel{IR}{=}\Lambda^4\frac{g^2c_2\theta\dim\mathfrak{g}}{2^4\pi^4}+\frac{g^2c_2\Lambda^2}{2^3\pi^2}\mathrm{Tr}\,[PS_{\mu\mu}(y,y)]\\
&-\frac{g^2c_2^2L^2W_{-1}}{2^7\pi^4}-\frac{5g^2c_2^2LW_{-1}}{3\pi^42^8}
+\frac{g^2c_2L}{2^4\pi^2}\mathrm{Tr}\,[F_{\nu\mu}(y)PS_{\nu\mu}(y,y)],
\end{align*}
\begin{align*}
I^6_{3}&=-\frac{g^2}{4}\int_{\mathbb{R}^4}d^4y\,f^{cab}f^{cde}G^{\,\,\,ad}_{1\mu\mu}(y,y)G^{\,\,\,be}_{1\nu\nu}(y,y)\\
&\stackrel{IR}{=}-\Lambda^4\frac{g^2c_2\theta\dim\mathfrak{g}}{2^2\pi^4}-\frac{g^2\Lambda^2c_2}{2\pi^2}\mathrm{Tr}\,[PS_{\mu\mu}(y,y)]+\frac{5g^2c_2^2LW_{-1}}{2^6\pi^43}.
\end{align*}
Therefore, the total contribution of the sixth diagram is equal to
\begin{equation}
\mathcal{J}_6-\mathcal{J}_6\big|_{B=0}\stackrel{IR}{=}\frac{5g^2c_2^2LW_{-1}}{2^8\pi^4}-\frac{3g^2c_2^2L^2W_{-1}}{2^7\pi^4}+\frac{3g^2c_2L}{2^4\pi^2}\mathrm{Tr}\,[F_{\nu\mu}(y)PS_{\nu\mu}(y,y)].
\end{equation}

\subsection{Other $\mathcal{J}$-components and sum of them all}
\label{oth}
All calculations for the components $\mathcal{J}_i$, i=2,\ldots,5, is produced in the Appendix B. So we can present only results
\begin{equation}
\mathcal{J}_2-\mathcal{J}_2\big|_{B=0}\stackrel{IR}{=}
\frac{g^2c_2^2S^3W_{-1}}{2^7\pi^6}2L^2-
\frac{g^2c_2S^3L}{2^3\pi^4}
\mathrm{Tr}\,[PS_{\alpha\beta}(y,y)F_{\alpha\beta}(y)],
\end{equation}
\begin{equation}
\mathcal{J}_3-\mathcal{J}_3\big|_{B=0}\stackrel{IR}{=}
\frac{g^2c_2^2S^3W_{-1}}{2^7\pi^6}\left(-\frac{L}{4}+\frac{L^2}{2}\right)-
\frac{g^2c_2S^3L}{2^5\pi^4}
\mathrm{Tr}\,[PS_{\alpha\beta}(y,y)F_{\alpha\beta}(y)],
\end{equation}
\begin{equation}
\mathcal{J}_4-\mathcal{J}_4\big|_{B=0}\stackrel{IR}{=}
\frac{g^2c_2^2S^3W_{-1}}{2^7\pi^6}\left(-\frac{L}{8}\right),
\end{equation}
\begin{equation}
\mathcal{J}_5-\mathcal{J}_5\big|_{B=0}\stackrel{IR}{=}
\frac{g^2c_2^2S^3W_{-1}}{2^7\pi^6}\left(\frac{L}{4}\right).
\end{equation}
Now we are ready to find the sum of all contributions. Using the previous formulas, we get
\begin{equation}
\label{eq57}
\sum_{i=1}^{6}(\mathcal{J}_i-\mathcal{J}_i\big|_{B=0})\stackrel{IR}{=}
\frac{g^2c_2^2W_{-1}}{(4\pi)^4}\left(\frac{9L}{2}\right).
\end{equation}
\textbf{Remark:} The last formula differs from the final answer (\ref{eq58}) by the factor $-1/4$, where the minus follows from the second line of the decomposition (\ref{eq20}), because $\mathcal{J}$-components do not include it, and the factor $1/4$ goes from the formula $4g^2=\alpha$.

\section{Discussion}
The result obtained above is the most general. In the sense that we did not use a gauge fixing procedure for the background field anywhere. At the same time, we know that the background field is a solution of the quantum equation of motion and hence has additional properties. Sometimes \cite{34} these properties are useful and can be used in searching for singularities to transform some coefficients into others. However, the Yang--Mills theory is designed in such a way that the quantum equation of motion is not used at this stage. This is probably due to the fact that the theory has only one dimensionless constant $\alpha$.\\

In the work we calculated the second $\beta$-function coefficient, and it has a different value from the coefficient calculated in the case of dimensional regularization \cite{13}. 
At first glance this is a bit unexpected, because the coefficients for the higher singularities should be the same regardless of the choice of regularization, but we believe this is consistent with the general construction. Indeed, one of the reasons may be the cutoff regularization, which can violate the gauge invariance. In this case it is possible to recover the invariance by a procedure described in \cite{w3,w4}. Similar situation was also discussed in \cite{w9,w10}.

\paragraph{Acknowledgments.}
This research is fully supported by the grant in the form of subsidies from the Federal budget for state support of creation and development world-class research centers, including international mathematical centers and world-class research centers that perform research and development on the priorities of scientific and technological development. The agreement is between MES and PDMI RAS from \textquotedblleft8\textquotedblright\,November 2019 № 075-15-2019-1620.\\

We are thankful to K.V. Stepanyants and A.L. Kataev for additional bibliography links. Also, A.V. Ivanov is a winner of the Young Russian Mathematician award and would like to thank its sponsors and jury.

\section{Appendix A} 
\label{apa}
By using definitions and notations described above (see the first part of Section \ref{s6}) we can formulate set of formulas:
\begin{equation}
\label{eq9}
(1,F_{\mu\nu},F_{\rho\nu})=-\frac{1}{2}(1,1,F_{\mu\nu}F_{\rho\nu})
=\frac{c^2_2}{2}\rho_{\mu\rho},
\end{equation}
\begin{equation}
\label{eq5}
(\Phi,\Phi, a_{1\beta\beta})=-\frac{c_2^2\varkappa}{3}+O(|x-y|^3),
\end{equation}
\begin{equation}
\label{eq10}
(x-y)^\alpha(\Phi,\Phi,\overrightarrow{D}_{x^\alpha} a_{1\beta\beta})=-\frac{2c_2^2\varkappa}{3}+O(|x-y|^3),
\end{equation}
\begin{equation}
\label{eq11}
(x-y)^\alpha(\Phi,\Phi,(a_{1\beta\beta}\overleftarrow{D}_{y^\alpha}))=\frac{2c_2^2\varkappa}{3}+O(|x-y|^3),
\end{equation}
\begin{equation}
\label{eq12}
\frac{(x-y)^{\nu\mu}}{r^2}(\Phi,\Phi,a_{1\nu\mu})=-\frac{c^2_2\varkappa}{12}+O(|x-y|^3),
\end{equation}
\begin{equation}
\label{eq13}
(x-y)^\nu(a_{1\mu\nu},\Phi,\overrightarrow{D}_{x^\mu}\Phi)=-\frac{c^2_2\varkappa}{2}+O(|x-y|^3),
\end{equation}
\begin{equation}
\label{eq14}
(x-y)^\mu(a_{1\mu\nu},\Phi,(\Phi\overleftarrow{D}_{y^\nu}))=\frac{c^2_2\varkappa}{2}+O(|x-y|^3),
\end{equation}
\begin{equation}
\label{eq15}
(x-y)^\nu(\Phi,\Phi,\overrightarrow{D}_{x^\mu}a_{1\nu\mu})=-\frac{7c^2_2\varkappa}{6}+O(|x-y|^3),
\end{equation}
\begin{equation}
\label{eq16}
(x-y)^\nu(\Phi,\Phi,(a_{1\nu\mu}\overleftarrow{D}_{y^\mu}))=-\frac{5c^2_2\varkappa}{6}+O(|x-y|^3),
\end{equation}
\begin{equation}
\label{eq17}
(\Phi,\Phi,a_{2\mu\rho})=2c^2_2\rho_{\mu\rho}-\frac{c_2^2}{12}\delta_{\mu\rho}\rho+O(|x-y|),
\end{equation}
\begin{equation}
\label{eq18}
(\Phi,\Phi,a_{2\beta\beta})=\frac{5}{3}c^2_2\rho+O(|x-y|),
\end{equation}
\begin{equation}
\label{eq19}
(\Phi,\Phi,a_2)=-\frac{c_2^2}{12}\rho+O(|x-y|),
\end{equation}
\begin{equation}
\label{eq56}
(\Phi,\Phi,\Phi)=c_2\dim\mathfrak{g}+O(|x-y|^5).
\end{equation}
Here we give a detailed proof only for the formula (\ref{eq5}). Other formulas are proved similarly by using the equalities from Section \ref{s4}. Let us use the Taylor expansions (\ref{eq6}) and (\ref{eq7}) for $\Phi(x,y)$ and $a_{1\beta\beta}(x,y)$, then we have
\begin{align*}
\label{a1}
(\Phi,\Phi,a_{1\beta\beta})&=(1,1,\frac{2}{3}(x-y)^{\sigma}\nabla_\rho F_{\sigma\rho})\\
&+2(1,-(x-y)^\alpha B_\alpha,\frac{2}{3}(x-y)^\sigma\nabla_\rho F_{\sigma\rho})\\
&+(1,1,\frac{1}{3}(x-y)^{\sigma_1\sigma_2}F_{\sigma_1\rho}F_{\sigma_2\rho})\\
&+(1,1,\frac{1}{6}(x-y)^{\sigma_1\sigma_2}\nabla_{(\rho}\nabla_{\sigma_1)}F_{\sigma_2\rho})\\
&-(1,1,\frac{2}{3}(x-y)^{\sigma_1\sigma_2}B_{\sigma_1}\nabla_\rho F_{\sigma_2\rho})+O(|x-y|^3).
\end{align*}
In the right hand side the first and the fourth terms are traceless, so they are equal to zero, and the second term cancels the fifth one due to the equality (\ref{eq8}). It means that we have only the second term which is equal to
\begin{equation*}
f^{aed}f^{ked}\left(\frac{(x-y)^{\sigma_1\sigma_2}}{3}F^c_{\sigma_1\rho}f^{ach}F^b_{\sigma_2\rho}f^{hbk}\right)=-\frac{c_2^2}{3}(x-y)^{\sigma_1\sigma_2}F^a_{\sigma_1\mu}F^a_{\sigma_2\mu},
\end{equation*}
where the relation (\ref{eq4}) was used twice.

\section{Appendix B}
\label{apb}
\paragraph{Contribution to $\mathcal{J}_2$-component:}
\begin{equation*}
I^2_{1,1,2}\stackrel{IR}{=}I^2_{1,2,1}\stackrel{IR}{=}\frac{g^2c^2_2S^3W_{-1}}{2^7\pi^6}\left(-\frac{L}{6}+\frac{L^2}{6}\right)
,\,\,\,\,\,\,
I^2_{2,1,1}\stackrel{IR}{=}\frac{g^2c^2_2S^3W_{-1}}{2^7\pi^6}\left(\frac{5L^2}{6}\right),
\end{equation*}
\begin{equation*}
I^2_{1,1,3}+I^2_{1,3,1}+I^2_{3,1,1}\stackrel{IR}{=}\frac{g^2c^2_2S^3W_{-1}}{2^7\pi^6}\left(-\frac{5L}{3}+\frac{5L^2}{6}\right),
\end{equation*}
\begin{equation*}
I^2_{1,2,2}+I^2_{2,1,2}+I^2_{2,2,1}\stackrel{IR}{=}\frac{g^2c^2_2S^3W_{-1}}{2^7\pi^6}L
,\,\,\,\,\,\,
I^2_{1,1,1}=-2I^1_{1,1,1},
\end{equation*}
\begin{equation*}
I^2_{1,1,4}+I^2_{1,4,1}\stackrel{IR}{=}
-\frac{g^2c_2S^3L}{2^4\pi^4}\mathrm{Tr}\,[\overrightarrow{D}_{x^\alpha}\overrightarrow{D}_{x^\alpha}PS_{\beta\beta}(x,y)],\,\,\,\,\,\,
I^2_{4,1,1}=-2I^1_{4,1,1},
\end{equation*}
\begin{equation*}
I^2_{4,1,2}\stackrel{IR}{=}I^2_{4,2,1}\stackrel{IR}{=}0.
\end{equation*}
So after summation we obtain:
\begin{equation}
\mathcal{J}_2^{loc}-\mathcal{J}_2^{loc}\big|_{B=0}
\stackrel{IR}{=}\frac{g^2c_2^2S^3W_{-1}}{2^7\pi^6}\left(-\frac{5L}{4}+2L^2\right),
\end{equation}
\begin{equation}
\mathcal{J}_2^{non-loc}\stackrel{IR}{=}
\frac{g^2c_2^2S^3W_{-1}}{2^7\pi^6}\left(\frac{5L}{4}\right)-
\frac{g^2c_2S^3L}{2^3\pi^4}
\mathrm{Tr}\,[PS_{\alpha\beta}(y,y)F_{\alpha\beta}(y)].
\end{equation}

\paragraph{Contribution to $\mathcal{J}_3$-component:}
\begin{equation*}
I^3_{1,1,2}=\frac{1}{2}I^2_{1,2,1},\,\,\,\,\,\,
I^3_{1,2,1}=I^1_{1,1,2}+I^2_{1,1,2},\,\,\,\,\,\,
I^3_{2,1,1}\stackrel{IR}{=}\frac{g^2c^2_2S^3W_{-1}}{2^7\pi^6}\left(-\frac{L^2}{12}\right),
\end{equation*}
\begin{equation*}
I^3_{1,1,3}+I^3_{1,3,1}+I^3_{3,1,1}\stackrel{IR}{=}\frac{g^2c^2_2S^3W_{-1}}{2^7\pi^6}\left(-\frac{5L}{6}+\frac{5L^2}{12}\right),
\end{equation*}
\begin{equation*}
I^3_{1,2,2}+I^3_{2,1,2}+I^3_{2,2,1}\stackrel{IR}{=}\frac{g^2c^2_2S^3W_{-1}}{2^7\pi^6}\left(\frac{L^2}{2}\right)
,\,\,\,\,\,\,
I^3_{1,1,1}=-I^1_{1,1,1},
\end{equation*}
\begin{equation*}
I^3_{1,1,4}\stackrel{IR}{=}-\frac{g^2c_2S^3L}{2^6\pi^4}\mathrm{Tr}\,[\overrightarrow{D}_{x^\alpha}\overrightarrow{D}_{x^\alpha}PS_{\beta\beta}(x,y)],\,\,\,\,\,\,
I^3_{1,4,1}\stackrel{IR}{=}-\frac{g^2c_2S^3L}{2^4\pi^4}\mathrm{Tr}\,[\overrightarrow{D}_{x^\mu}\overrightarrow{D}_{x^\nu}PS_{\nu\mu}(x,y)],
\end{equation*}
\begin{equation*}
I^3_{4,1,1}\stackrel{IR}{=}\frac{g^2c_2S^3\Lambda^2}{2^4\pi^4}\mathrm{Tr}\,[PS_{\mu\mu}(y,y)]
+\frac{g^2c_2S^3L}{2^6\pi^4}\mathrm{Tr}\,[\overrightarrow{D}_{x^\nu}\overrightarrow{D}_{x^\nu}PS_{\mu\mu}(x,y)],
\end{equation*}
\begin{equation*}
I^3_{4,1,2}\stackrel{IR}{=}-\frac{g^2c_2S^3L}{2^5\pi^4}\mathrm{Tr}\,[PS_{\mu\alpha}(x,y)F_{\mu\alpha}(x)],\,\,\,\,\,\,
I^3_{4,2,1}\stackrel{IR}{=}0.
\end{equation*}
So after summation we obtain:
\begin{equation}
\mathcal{J}_3^{loc}-\mathcal{J}_3^{loc}\big|_{B=0}
\stackrel{IR}{=}\frac{g^2c_2^2S^3W_{-1}}{2^7\pi^6}\left(-\frac{7L}{8}+\frac{L^2}{2}\right), 
\end{equation}
\begin{equation}
\mathcal{J}_3^{non-loc}\stackrel{IR}{=}
\frac{g^2c_2^2S^3W_{-1}}{2^7\pi^6}\left(\frac{5L}{8}\right)-
\frac{g^2c_2S^3L}{2^5\pi^4}
\mathrm{Tr}\,[PS_{\alpha\beta}(y,y)F_{\alpha\beta}(y)].
\end{equation}

\paragraph{Contribution to $\mathcal{J}_4$-component:}
\begin{equation*}
I^4_{1,1,2}=-\frac{1}{4}I^1_{1,2,1},\,\,\,\,\,\,
I^4_{1,2,1}=\frac{1}{4}I^3_{1,2,1},\,\,\,\,\,\,
I^4_{2,1,1}=\frac{1}{4}I^1_{2,1,1}+\frac{1}{2}I^3_{2,1,1},
\end{equation*}
\begin{equation*}
I^4_{1,1,3}+I^4_{1,3,1}+I^4_{3,1,1}\stackrel{IR}{=}\frac{g^2c^2_2S^3W_{-1}}{2^7\pi^6}\left(-\frac{5L}{24}+\frac{5L^2}{48}\right),
\end{equation*}
\begin{equation*}
I^4_{1,2,2}+I^4_{2,1,2}+I^4_{2,2,1}\stackrel{IR}{=}\frac{g^2c^2_2S^3W_{-1}}{2^7\pi^6}
\left(-\frac{L}{8}+\frac{L^2}{4}\right),\,\,\,\,\,\,I^4_{1,1,1}=
-\frac{1}{4}I^1_{1,1,1},
\end{equation*}
\begin{equation*}
I^4_{1,1,4}=-\frac{1}{4}I^1_{1,4,1},\,\,\,\,\,\,
I^4_{1,4,1}=\frac{1}{4}I^3_{1,4,1},
\end{equation*}
\begin{equation*}
I^4_{4,1,1}\stackrel{IR}{=}-\frac{1}{4}I^1_{4,1,1}+\frac{g^2c_2S^3L}{2^6\pi^4}\mathrm{Tr}\,[PS_{\nu\mu}(x,y)F_{\nu\mu}(x)],
\end{equation*}
\begin{equation*}
I^4_{4,1,2}+I^4_{4,2,1}\stackrel{IR}{=}-\frac{g^2c_2S^3L}{2^6\pi^4}\mathrm{Tr}\,[PS_{\nu\mu}(x,y)F_{\nu\mu}(y)].
\end{equation*}
So after summation we obtain:
\begin{equation}
\mathcal{J}_4^{loc}-\mathcal{J}_4^{loc}\big|_{B=0}
\stackrel{IR}{=}\frac{g^2c_2^2S^3W_{-1}}{2^7\pi^6}\left(-\frac{9L}{32}\right),
\end{equation}
\begin{equation}
\mathcal{J}_4^{non-loc}\stackrel{IR}{=}
\frac{g^2c_2^2S^3W_{-1}}{2^7\pi^6}\left(\frac{5L}{32}\right).
\end{equation}

\paragraph{Contribution to $\mathcal{J}_5$-component:}
\begin{equation*}
I^5_{1,1,2}\stackrel{IR}{=}I^5_{1,2,1}\stackrel{IR}{=}\frac{g^2c^2_2S^3W_{-1}}{2^7\pi^6}\left(\frac{L}{48}-\frac{L^2}{48}\right)
,\,\,\,\,\,\,
I^5_{2,1,1}\stackrel{IR}{=}\frac{g^2c^2_2S^3W_{-1}}{2^7\pi^6}\left(-\frac{5L^2}{48}\right),
\end{equation*}
\begin{equation*}
I^5_{1,1,3}+I^5_{1,3,1}+I^5_{3,1,1}\stackrel{IR}{=}\frac{g^2c^2_2S^3W_{-1}}{2^7\pi^6}\left(-\frac{L}{24}+\frac{7L^2}{48}\right),
\end{equation*}
\begin{equation*}
I^5_{1,2,2}\stackrel{IR}{=}I^5_{2,1,2}\stackrel{IR}{=}I^5_{2,2,1}\stackrel{IR}{=}0,\,\,\,\,\,\,
I^5_{1,1,1}\stackrel{IR}{=}\frac{1}{4}I^1_{1,1,1},
\end{equation*}
\begin{equation*}
I^5_{1,1,4}\stackrel{IR}{=}
I^5_{1,4,1}\stackrel{IR}{=}\frac{g^2c_2S^3L}{2^6\pi^4}\mathrm{Tr}\,[\overrightarrow{D}_{x^{\nu}}\overrightarrow{D}_{x^{\nu}}PS(x,y)],
\end{equation*}
\begin{equation*}
I^5_{4,1,1}\stackrel{IR}{=}\frac{1}{4}I^1_{4,1,1},
\,\,\,\,\,\,
I^5_{4,1,2}\stackrel{IR}{=}I^5_{4,2,1}\stackrel{IR}{=}0.
\end{equation*}
So after summation we obtain:
\begin{equation}
\mathcal{J}_5^{loc}-\mathcal{J}_5^{loc}\big|_{B=0}
\stackrel{IR}{=}\frac{g^2c_2^2S^3W_{-1}}{2^7\pi^6}\left(\frac{L}{32}\right),
\end{equation}
\begin{equation}
\mathcal{J}_5^{non-loc}\stackrel{IR}{=}
\frac{g^2c_2^2S^3W_{-1}}{2^7\pi^6}\left(\frac{7L}{32}\right).
\end{equation}

\section{Appendix C}
In the section we want to discuss an ansatz for the Green function regular part in the four-dimensional case. Let $\mathcal{A}=-\mathbb{1}\overrightarrow{D}_{\mu}\overrightarrow{D}_{\mu}-v$ be a Laplace-type operator with the covariant derivative (\ref{eq50}) and an arbitrary potential $v$. As it was noted above (see \cite{29}) we can represent a solution of the equation
\begin{equation}
\label{eq51}
\mathcal{A}\,\mathcal{G}(x,y)=\mathbb{1}\delta(x-y)
\end{equation}
in the form
\begin{equation*}
\frac{\mathfrak{a}_{0}(x,y)}{4\pi^2r^2}+
\sum_{k=1}^{\infty}(-1)^k\frac{r^{2k-2}\ln(r^2\mu^2)}{(k-1)!4^{k+1}\pi^2}\mathfrak{a}_{k}(x,y)+
\mathcal{PS}(x,y),
\end{equation*}
where the regular part $\mathcal{PS}$ may depend on $\mu$. Then, using the equality (\ref{eq51}) and relations for the Seeley--DeWitt coefficients \cite{33}, we get the equation for the regular part
\begin{equation}
\label{eq53}
(\mathcal{A}\,\mathcal{PS})(x,y)=f(x,y),
\end{equation}
where the last function is regular and does not depend on $\mu$. Now we use a dimensional analysis to find a series for $\mathcal{PS}$. Let $l$ denotes a dimension of length, so we have
\begin{equation*}
[\mathcal{PS}]\sim l^{-2},\,\,\,\,\,\,[(x-y)^\nu]\sim l,\,\,\,\,\,\,[v]\sim l^{-2},
\end{equation*}
\begin{equation*}
[\nabla_{\mu}]\sim l^{-1},\,\,\,\,\,\,[F_{\mu\nu}]\sim l^{-2},\,\,\,\,\,\,[\mu]\sim l^{-1}.
\end{equation*}
Thereby Taylor coefficients of $\mathcal{PS}(x,y)$ near the point $y$ are constructed by the following blocks:
\begin{equation*}
\nabla_{\mu_1}\ldots\nabla_{\mu_i}F_{\mu\nu},\,\,\,\,\,\,
\nabla_{\mu_1}\ldots\nabla_{\mu_j}v,\,\,\,\,\,\,
\mu^{-k}.
\end{equation*}
This means that the potential index and the Lorentzian one cannot be convoluted, even if their dimensions are the same. From this observation the ansatz (\ref{eq52}) follows. Let us show that, actually, factors $\mu^{-k}$ do not appear. Otherwise we could have a decomposition
\begin{equation*}
\mathcal{PS}=\sum_{k=0}^{\infty}\mu^{-k}\mathcal{PS}_k.
\end{equation*}
So from the formula (\ref{eq53}) we see, that $\mathcal{PS}_k$, where $k>0$, are zero modes of the operator $\mathcal{A}$ that are cancelled by (\ref{eq21}).

\end{document}